\documentclass[a4paper,12pt]{revtex4}

\usepackage{comment}
\usepackage{amssymb}
\usepackage{color}
\usepackage{amsmath}
\usepackage{ulem}
\usepackage{graphicx}
\usepackage{dcolumn}
\usepackage{bm}
\usepackage{slashed}
\usepackage{indentfirst}
\usepackage{booktabs}
\usepackage{amssymb,amsfonts}
\usepackage{amsmath}
\usepackage{color}

\begin{document}

\title{Strong decay modes $\bar{K}\Xi$ and $\bar{K}\Xi\pi$ of the  $\Omega(2012)$ in the $\bar{K}\Xi(1530)$ and $\eta\Omega$ molecular scenario}

\author{Yin Huang}
\affiliation{School of Physics and Nuclear Energy Engineering, Beihang University, Beijing 100191, China}

\author{Ming-Zhu Liu}
\affiliation{School of Physics and Nuclear Energy Engineering, Beihang University, Beijing 100191, China}
\author{Jun-Xu Lu}
\affiliation{School of Physics and Nuclear Energy Engineering, Beihang University, Beijing 100191, China}

\author{Ju-Jun Xie} \email{xiejujun@impcas.ac.cn}

\affiliation{Institute of Modern Physics, Chinese Academy of
Sciences, Lanzhou 730000, China}

\author{Li-Sheng Geng} \email{lisheng.geng@buaa.edu.cn}
\affiliation{School of Physics and Nuclear Energy Engineering,
International Research Center for Nuclei and Particles in the Cosmos
and Beijing Key Laboratory of Advanced Nuclear Materials and
Physics, Beihang University, Beijing 100191, China}

\begin{abstract}
We study the $\bar{K} \Xi$ decay mode of the newly observed $\Omega(2012)$ assuming  that the $\Omega(2012)$ is a dynamically generated state with spin-parity $J^P = 3/2^-$
from the coupled channel $S$-wave interactions of $\bar{K}\Xi(1530)$ and $\eta \Omega$. In addition we also calculate its $K\pi\Xi$ three-body decay mode.
It is shown that the so-obtained total decay width  is in fair agreement with the experimental data. We compare our results with those of other recent studies and highlight differences
among them.
\end{abstract}

\maketitle

\section{Introduction}

Very recently, the Belle Collaboration observed an $\Omega$ excited state in the $\Xi^0 K^-$ and $\Xi^- K^0_S$ invariant mass distributions~\cite{Yelton:2018mag}. Its  mass and width are determined to be $M = 2012.4 \pm 0.7 \pm 0.6$ MeV and $\Gamma = 6.4^{+2.5}_{-2.0}  \pm 1.6$ MeV. The existence of such $\Omega$ excited states with a mass around 2000 MeV has already been predicted by various models, such as quenched quark models~\cite{Capstick:1986bm,Loring:2001ky,Pervin:2007wa,Faustov:2015eba}, the Skyrme model~\cite{Oh:2007cr}, and lattice gauge theory~\cite{Engel:2013ig}. On the other hand, the extended quark models~\cite{Yuan:2012zs,An:2013zoa,An:2014lga}, where the instanton-induced quark-antiquark pair creation or Nambu-Jona-Lasino interaction was employed, predicted  $\Omega$ states with negative parity but lower masses, the lowest $\Omega$ state lying around 1800 MeV, about 200 MeV lower than those predicted in Refs.~\cite{Capstick:1986bm,Loring:2001ky,Pervin:2007wa,Faustov:2015eba,Oh:2007cr,Engel:2013ig}. One of the reasons is that the $\Omega$ states in Refs.~\cite{Yuan:2012zs,An:2013zoa,An:2014lga} have large five-quark components,

In Refs.~\cite{Kolomeitsev:2003kt,Sarkar:2004jh,Si-Qi:2016gmh}, the interactions of the $\bar{K}\Xi(1530)$ and $\eta \Omega$ coupled channels were investigated in the chiral unitary approach. An $\Omega$ excited state with a mass around 2012 MeV and  $J^P =3/2^-$ can be dynamically generated  by use of a reasonable subtraction constant.

After the observation of the $\Omega(2012)$, its two-body strong decays were studied within the chiral quark model in Ref.~\cite{Xiao:2018pwe}, where it was shown that the newly observed $\Omega(2012)$ could be assigned to the $J^P = 3/2^-$ three quark state.  In Ref.~\cite{Aliev:2018syi} the mass and residue of the $\Omega(2012)$ were calculated by employing the QCD sum rule method with the conclusion that the $\Omega(2012)$ could be a $1P$ orbitally excited $\Omega$ state with $J^P =3/2^-$.  The analysis of Ref.~\cite{Aliev:2018syi} was extended in Ref.~\cite{Aliev:2018} to study the $\Omega(2012) \to K^- \Xi^0$ decay. In Ref.~\cite{Polyakov:2018mow},  the authors performed a  flavor $SU(3)$ analysis  and concluded that the preferred $J^P$ for the $\Omega(2012)$ is also $3/2^-$. In Refs.~\cite{Valderrama:2018bmv,Lin:2018nqd}, its strong decay modes were studied assuming that the $\Omega(2012)$ is a $\bar{K}\Xi(1530)$ hadronic molecule.

In this work, we take the chiral unitary approach and assume that the $\Omega(2012)$ is a dynamically generated state from the $\bar{K} \Xi(1530)$ and $\eta \Omega$ interactions. The coupling of the $\Omega(2012)$ to $\bar{K}\Xi(1530)$ is then obtained from the residule at the pole position. We then
calculate its decay into $K \Xi$ via a triangle diagram. We also calculate the three-body partial decay width of the $\Omega(2012)$ into $K\Xi\pi$.
 The total decay width compares favorably  with the experimental data~\cite{Yelton:2018mag} and agrees with other theoretical approaches.

 This work is organized as follows. In Section II, we briefly explain the chiral unitary approach and calculate the two and three body decays of the $\Omega(2012)$. Results
 and discussions are shown in Section III, followed by a short summary in Section IV.

 \section{Formalism}

\subsection{$\Omega(2012)$ as a $\bar{K}\Xi(1530)$ and $\eta \Omega$ molecula}

The mass of the $\Omega(2012)$ is slightly below the $\bar{K} \Xi(1530)$ threshold. It is natural to treat it as a $\bar{K} \Xi(1530)$ molecular state, dynamically generated from the interaction of the coupled channels $\bar{K}\Xi(1530)$ and $\eta\Omega$ in the isospin $I=0$ sector. However, the possibility to be an $I=1$ molecule  cannot be excluded~\cite{Polyakov:2018mow}. Within the chiral unitary approach, the interaction of the coupled channels $\bar{K} \Xi(1530)$ and $\eta \Omega$ in the strangeness $-3$ and isospin 0 was first studied in Ref.~\cite{Sarkar:2004jh}, where a pole at $(2141-i38)$ MeV was found with a natural subtraction constant $a_{\mu} = -2$ and a renormalization scale $\mu = 700$ MeV. Later,  it was explicitly shown  in Ref.~\cite{Si-Qi:2016gmh} that the pole position of the $3/2^-$ $\Omega$ state can shift by varying $a_{\mu}$. If we take $a_{\mu} = -2.5$ and $\mu=700$ MeV, we obtain a pole at $z_R = (2012.7, i0)$ MeV, which can be associated to the newly observed $\Omega$(2012) state~\cite{Yelton:2018mag}.  In the cutoff regularization scheme, the
corresponding momentum cutoff is $\Lambda= $726 MeV, which seems to quite natural as well.

The couplings of the bound state to the coupled  channels, $\bar{K}\Xi(1530)$ (channel 1) and $\eta \Omega$ (channel 2), can be obtained from the residue of the
scattering amplitude at the pole position  $z_R$,
\begin{eqnarray}
T_{ij} = \frac{g_{ii}g_{jj}}{\sqrt{s} - z_R},
\end{eqnarray}
where $g_{ii}$ is the coupling of the state to the $i$-th channel. One finds
\begin{eqnarray}
g_{11} = 1.65, ~~~~~~~~~~g_{22} = 2.80.
\end{eqnarray}

Then one can write down the effective interaction of $\Omega(2012)\bar{K}\Xi(1530)$ ($\equiv \Omega^* \bar{K} \Xi^*$) and $\Omega(2012)\eta \Omega$ ($\equiv \Omega^* \eta \Omega$),
\begin{eqnarray}
v_{\Omega^* \bar{K} \Xi^*} &=& \frac{g_{11}}{\sqrt{2}} \bar{\Xi}^{* \mu} \Omega^*_{\mu} \phi_{\bar{K}}, \\
v_{\Omega^* \eta \Omega} &=& g_{22} \bar{\Omega}^{\mu} \Omega^*_{\mu} \phi _\eta.
\end{eqnarray}

It is worth to mention that the $g_{\Omega^*(2012)\bar{K}\Xi(1530)}$  obtained in Ref.~\cite{Lin:2018nqd} with the assumption that the $\Omega^*(2012)$ is a pure $S$-wave $\bar{K} \Xi(1530)$ hadronic molecule with spin-parity $3/2^-$ is about 1.97, which is different from ours  by a mere $16\%$, and the difference is only $2.6\%$ in terms of $g^2_{\Omega^*(2012)\bar{K}\Xi(1530)}$. For the $\Omega^*(2012) \to \bar{K} \pi \Xi$ three-body decay, since only the $\bar{K} \Xi(1530)$ component contributes at  tree level and the partial decay width is proportional to $g^2_{\Omega^*(2012)\bar{K}\Xi(1530)}$, our three-body decay width is almost the same as that of Ref.~\cite{Lin:2018nqd}.

\subsection{The $\Omega(2012) \to \bar{K} \Xi$ and $\bar{K}\Xi\pi$ decays}

\begin{figure}[htbp]
\begin{center}
\includegraphics[scale=0.45]{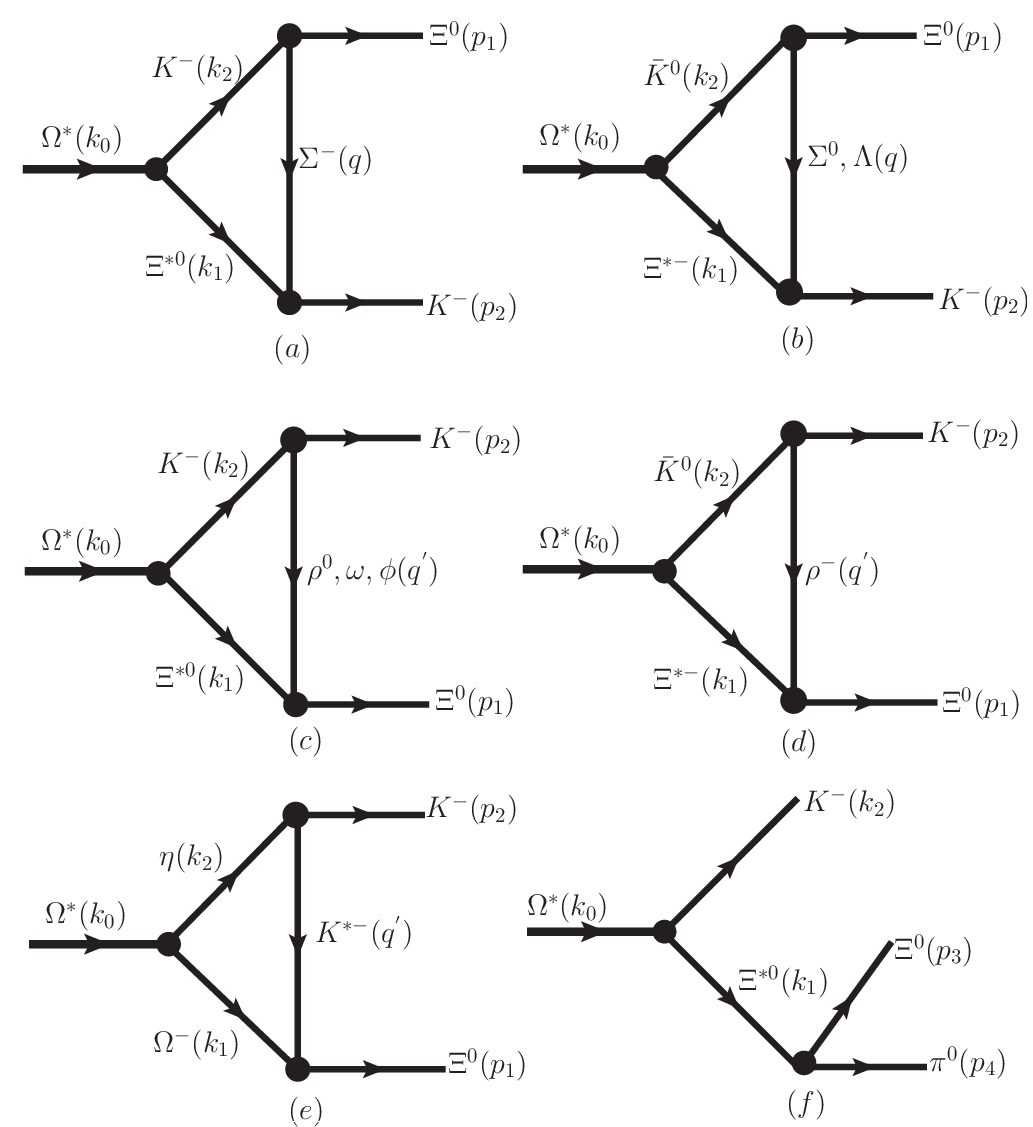}
\caption{Feynman diagrams contributing to the decays of the $\Omega(2012)$  to $\bar{K}\Xi$ and $\bar{K}\Xi\pi$.}\label{mku}
\end{center}
\end{figure}

In the present work, we assume that the $\Omega(2012)$ ($\equiv \Omega^{*-}$) exists  and has a mass as that reported by the Belle Collaboration, and study its strong decays to the two-body final state $\bar{K}\Xi$ assuming  that it is a molecular state of $\bar{K}\Xi(1530)$ and $\eta\Omega$, as predicted by the chiral unitary approach~\cite{Sarkar:2004jh}. Then, the $\Omega(2012) \to \bar{K} \Xi$ decay can proceed through the triangular diagrams as shown in Fig.~\ref{mku}[(a)-(e)], where the $\Sigma$, $\Lambda$, $\rho$, $\omega$, $\phi$, and $K^{*}$ exchanges are considered. Note that by considering those diagrams of Fig.~\ref{mku} we can easily show that the partial decay widths of $\Omega^{*-} \to K^- \Xi^0$ and $\Omega^{*-} \to K^0 \Xi^-$ are the same and we will explicitly calculate the decay of $\Omega^{*-} \to K^- \Xi^0$ in the following.   Compared with the decays to the two final states, the contribution to the three-body decay $\Omega(2012) \to \bar{K} \Xi \pi$ only comes from the $\bar{K}\Xi(1530)$ component.  The decay width of the $\Xi(1530)$ listed in PDG is around 9 MeV and the partial decay width $\Xi(1530)\to\Xi\pi$ is the largest and almost saturates its total decay width.  Therefore, we only compute the three-body decay through the decay of the $\Xi(1530)$ and the simplest Feynman diagram is shown in Fig.~\ref{mku}[(f)].   Considering the quantum numbers and phase space, in addition to
the final state $K^{-}\Xi^{0}\pi^{0}$ shown in Fig.~\ref{mku}[(f)], the final states $\bar{K}^{0}\Xi^{-}\pi^{0}$, $\bar{K}^{0}\Xi^{0}\pi^{-}$, and  $K^{-}\Xi^{-}\pi^{+}$  should also be calculated.

In order to evaluate the decay amplitudes of the diagrams shown in Fig.~\ref{mku}, we need  the following effective Lagrangians to calculate the relevant vertices~\cite{Xie:2010zza,Nakayama:2006ty,Huang:2018wgr},
\begin{align}
&{\cal{L}}_{\Xi\Lambda{}K}=\frac{ig_{\Xi{}\Lambda{}K}}{m_{\Lambda}+m_{\Xi}}\partial^{\mu}\bar{K}\bar{\Lambda}\gamma_{\mu}\gamma_5\Xi+H.c.,\\
&{\cal{L}}_{\Xi\Sigma{}K}=\frac{ig_{\Xi{}\Sigma{}K}}{m_{\Sigma}+m_{\Xi}}\partial^{\mu}\bar{K}\bar{\Sigma}\cdot\tau\gamma_{\mu}\gamma_5\Xi+H.c.,\\
&{\cal{L}}_{\Xi^{*}\Sigma{}K}=\frac{f_{\Xi^{*}\Sigma{}K}}{m_{K}}\partial^{\nu}\bar{K}\bar{\Xi}^{*}_{\nu}\gamma_5\tau\cdot\Sigma+H.c.,\\
&{\cal{L}}_{\Xi^{*}\Lambda{}K}=\frac{f_{\Xi^{*}\Lambda{}K}}{m_{K}}\partial^{\nu}\bar{K}\bar{\Xi}^{*}_{\nu}\gamma_5\Lambda+H.c.,\\
&{\cal{L}}_{KK\rho}=ig_{\rho{}KK}[\bar{K}\partial_{\mu}K-\partial_{\mu}\bar{K}K]\vec{\tau}\cdot{}\vec{\rho}^{\mu},\\
&{\cal{L}}_{KK\omega}=ig_{\omega{}KK}[\bar{K}\partial_{\mu}K-\partial_{\mu}\bar{K}K]\omega^{\mu},\\
&{\cal{L}}_{KK\phi}=ig_{\phi{}KK}[\bar{K}\partial_{\mu}K-\partial_{\mu}\bar{K}K]\phi^{\mu},\\
&{\cal{L}}_{\eta{}KK^{*}}=ig_{\eta{}KK^{*}}[\bar{K}\partial_{\mu}\eta-\partial_{\mu}\bar{K}\eta]K^{*\mu},\\
&{\cal{L}}_{\rho\Xi\Xi^{*}}=i\frac{g_{\rho\Xi\Xi^{*}}}{m_{\rho}}\bar{\Xi}^{*\mu}\gamma^{\nu}\gamma_5[\partial_{\mu}\vec{\tau}\cdot\vec{\rho}_{\nu}-\partial_{\nu}\vec{\tau}\cdot\vec{\rho}_{\mu}]\Xi+H.c.,\\
&{\cal{L}}_{\omega\Xi\Xi^{*}}=i\frac{g_{\omega\Xi\Xi^{*}}}{m_{\omega}}\bar{\Xi}^{*\mu}\gamma^{\nu}\gamma_5[\partial_{\mu}\omega_{\nu}-\partial_{\nu}\omega_{\mu}]\Xi+H.c.,\\
&{\cal{L}}_{\phi\Xi\Xi^{*}}=i\frac{g_{\phi\Xi\Xi^{*}}}{m_{\phi}}\bar{\Xi}^{*\mu}\gamma^{\nu}\gamma_5[\partial_{\mu}\phi_{\nu}-\partial_{\nu}\phi_{\mu}]\Xi+H.c.,\\
&{\cal{L}}_{K^{*}\Xi\Omega}=i\frac{g_{K^{*}\Xi\Omega}}{m_{K^{*}}}\bar{\Omega}^{\mu}\gamma^{\nu}\gamma_5[\partial_{\mu}K^{*}_{\nu}-\partial_{\nu}K^{*}_{\mu}]\Xi+H.c.,\\
&{\cal{L}}_{\pi\Xi\Xi^{*}}=\frac{g_{\pi\Xi\Xi^{*}}}{m_{\pi}}\bar{\Xi}\partial^{\mu}\vec{\tau}\cdot\vec{\pi}\Xi^{*}_{\mu}+H.c\label{eq33}.
\end{align}
Within the $SU(3)$ symmetry, we determine the coupling constants to be $g_{\eta\bar{K}\bar{K}^{*}}=\sqrt{3}g_{\pi\bar{K}\bar{K}^{*}}=5.23$~\cite{Ronchen:2012eg}, $g_{\phi\Xi\Xi^{*}}=g_{\rho\Xi\Xi^{*}}=g_{\omega\Xi\Xi^{*}}=\frac{1}{\sqrt{6}}g_{\rho\Delta{}N}=-2.48$~\cite{Gao:2010ve}, and $g_{K^{*}\Xi\Omega}=-7.01$~\cite{Gao:2010ve}.  The coupling constant $g_{\pi\Xi\Xi^{*}}=2.24$($g_{\pi\Xi\Xi^{*}}=2.04$) is evaluated using Eq.~(\ref{eq33}) and the partial decay width $\Gamma_{\Xi^{*}\to\Xi\pi}=\Gamma_{\Xi^{*}}=9.1$ MeV($\Gamma_{\Xi^{*}\to\Xi\pi}=\Gamma_{\Xi^{*}}=9.5$ MeV) in the $\Xi^{*}$ rest frame. The masses of the particles needed in the present work are listed in Table~\ref{table-mass}.
The other coupling
constants are taken from Refs.~\cite{Xie:2010zza,Nakayama:2006ty,Huang:2018wgr} and are listed in Table~\ref{table0}.

\begin{table}[h!]
\centering
\caption{Masses of the particles needed in the present work (in units of MeV).}\label{table-mass}
\begin{tabular}{ccccccc}
\hline\hline
$\pi^{\pm}$~& $\pi^{0}$   & ~$\eta$   & ~$\rho$      &~$\omega$          & ~$\phi$           & ~$K^{\pm}$                                    \\ \hline
139.57     ~&  134.98     & ~547.86   & ~775.26      &~782.65            & ~1019.46          & ~493.68                                               \\  \hline \hline
$K^{0}$    ~& $K^{*}$     & ~$\Xi^{0}$& ~$\Xi^{-}$   & ~$\Xi^{0}(1530)$  & ~$\Xi^{-}(1530)$               \\ \hline
497.61     ~& 891.76      & ~1314.86  & ~1321.71     & ~1531.80          & ~1535.00                       \\  \hline \hline
\end{tabular}
\end{table}

\begin{table}[h!]
\centering
\caption{Values of the effective meson-baryon  and meson-meson coupling constants.}\label{table0}
\begin{tabular}{ccccccc}
\hline\hline
$g_{KK\phi}$&~~~~~~$g_{KK\rho}$ & ~~~~~~$g_{KK\omega}$ & ~~~~~~$g_{\Xi^{*}\Sigma{}K}$ &~~~~~~$g_{\Xi^{*}\Lambda{}K}$& ~~~~~~$g_{\Xi\Sigma{}K}$       & ~~~~~~$g_{\Xi\Lambda{}K}$                                   \\ \hline
-3.02    &~~~~~~  -3.02        & ~~~~~-3.02           & ~~~~~~3.22                   &~~~~~~5.58                   & ~~~~~~-13.26                   & ~~~~~~3.37                                                 \\  \hline \hline
\end{tabular}
\end{table}

With above effective interaction Lagrangians and the coupling constants, we obtain the decay amplitudes for $\Omega^{*}(2012)\to{}K^{-}\Xi^{0}$ and $\Omega^{*}(2012)\to{}\bar{K}\Xi\pi$, corresponding to the diagrams shown in Fig.~\ref{mku},

\begin{align}
{\cal{M}}_{\Sigma^{-}}&=\frac{\sqrt{2}g_{11}g_{\Xi^{0}\Sigma^{-}K^{-}}g_{\Xi^{*0}\Sigma^{-}K^{-}}}{m_{K^{-}}(m_{\Sigma^{-}}+m_{\Xi^{0}})}\int{}\frac{d^4q}{(2\pi)^4}\bar{u}(p_1)k\!\!\!/_2\gamma_{5}\frac{1}{q\!\!\!/-m_{\Sigma^{-}}+i\epsilon}\nonumber\\
                      &\times\gamma_5{\cal{P}}^{\nu\eta}u^{\eta}_{\Omega^{*}}(k_0)p_2^{\nu}\frac{1}{k_2^2-m^2_{K^{-}}+i\epsilon},\label{eq1}
\end{align}
\begin{align}
{\cal{M}}_{\Sigma^{0}}&=\frac{g_{11}g_{\Xi^{0}\Sigma^{0}\bar{K}^{0}}g_{\Xi^{*-}\Sigma^{0}K^{-}}}{\sqrt{2}m_{K^{-}}(m_{\Sigma^{0}}+m_{\Xi^{-}})}\int{}\frac{d^4q}{(2\pi)^4}\bar{u}(p_1)k\!\!\!/_2\gamma_{5}\frac{1}{q\!\!\!/-m_{\Sigma^{0}}+i\epsilon}\nonumber\\
                      &\times{}\gamma_5{\cal{P}}^{\nu\eta}u^{\eta}_{\Omega^{*}}(k_0)p_2^{\nu}\frac{1}{k_2^2-m^2_{K^{-}}+i\epsilon},\label{eq2}
\end{align}
\begin{align}
{\cal{M}}_{\Lambda}&=\frac{g_{11}g_{\Xi^{0}\Lambda\bar{K}^{0}}g_{\Xi^{*-}\Lambda{}K^{-}}}{\sqrt{2}m_{K^{-}}(m_{\Lambda}+m_{\Xi^{-}})}\int{}\frac{d^4q}{(2\pi)^4}\bar{u}(p_1)k\!\!\!/_2\gamma_{5}\frac{1}{q\!\!\!/-m_{\Lambda}+i\epsilon}\nonumber\\
                      &\times{}\gamma_5{\cal{P}}^{\nu\eta}u^{\eta}_{\Omega^{*}}(k_0)p_2^{\nu}\frac{1}{k_2^2-m^2_{K^{-}}+i\epsilon},\label{eq3}
\end{align}
\begin{align}
{\cal{M}}_{\rho^{0}/\omega/\phi}&=\frac{-ig_{11}g_{\Xi^{*}\Xi\rho/\omega/\phi}g_{\rho/\omega/{\phi}\bar{K}\bar{K}}}{\sqrt{2}m_{\rho/\omega}}\int{}\frac{d^4q^{'}}{(2\pi)^4}\bar{u}(p_1)\gamma^{\mu}\gamma_5\nonumber\\
                           &\times{}{\cal{P}}^{\nu\eta}u^{\eta}_{\Omega^{*}}(k_0)(q^{'}_{\nu}g_{\mu\alpha}-q^{'}_{\mu}g_{\nu\alpha})(k_1^{\beta}-p_2^{\beta})\nonumber\\
                           &\times\frac{1}{k_2^2-m^2_{K^{-}}+i\epsilon}\frac{-g^{\alpha\beta}+q^{'\alpha}q^{'\beta}/m^2_{\rho/\omega}}{q^{'2}-m^2_{\rho/\omega/\phi}+im_{\rho/\omega/\phi}\Gamma_{\rho/\omega/\phi}},\label{eq4}
\end{align}

\begin{align}
{\cal{M}}_{\rho^{-}}&=\frac{-i\sqrt{2}g_{11}g_{\Xi^{*}\Xi\rho}g_{\rho{}\bar{K}\bar{K}}}{m_{\rho}}\int{}\frac{d^4q^{'}}{(2\pi)^4}\bar{u}(p_1)\gamma^{\mu}\gamma_5\nonumber\\
                    &\times{}{\cal{P}}^{\nu\eta}u^{\eta}_{\Omega^{*}}(k_0)(q^{'}_{\nu}g_{\mu\alpha}-q^{'}_{\mu}g_{\nu\alpha})(k_1^{\beta}-p_2^{\beta})\nonumber\\
                    &\times\frac{1}{k_2^2-m^2_{K^{-}}+i\epsilon}\frac{-g^{\alpha\beta}+q^{'\alpha}q^{'\beta}/m^2_{\rho}}{q^{'2}-m^2_{\rho}+im_\rho\Gamma_{\rho}},\label{eq5}
\end{align}

\begin{align}
{\cal{M}}_{K^{*-}}&=\frac{-i g_{22}g_{\eta{K^{-}}K^{*-}}g_{\Omega^{-}K^{*-}\Xi^{0}}}{m_{K^{*-}}}\int{}\frac{d^4q^{'}}{(2\pi)^4}\bar{u}(p_1)\gamma^{\mu}\gamma_5\nonumber\\
                    &\times{}{\cal{P}}^{\nu\eta}u^{\eta}_{\Omega^{*}}(k_0)(q^{'}_{\nu}g_{\mu\alpha}-q^{'}_{\mu}g_{\nu\alpha})(k_1^{\beta}-p_2^{\beta})\nonumber\\
                    &\times\frac{1}{k_2^2-m^2_{\eta}+im_{\eta}\Gamma_{\eta}}\frac{-g^{\alpha\beta}+q^{'\alpha}q^{'\beta}/m^2_{K^{*}}}{q^{'2}-m^2_{K^{*}}+im_{K^*}\Gamma_{K^*}}\label{eq6},
\end{align}

\begin{align}
{\cal{M}}&_{\bar{K}(k_2)\Xi(p_3)\pi(p_4)}=f_I\frac{ig_{\pi\Xi\Xi^{*}}g_{\Omega^{*}\bar{K}\Xi^{*}}}{m_{\pi}}F(k^2_1)\bar{u}(p_3)p_3^{\nu}{\cal{P}}^{\nu\eta}u_{\eta}(k_0),
\end{align}
where
\begin{align} {\cal{P}}^{\nu\eta}=\frac{k\!\!\!/_1+m_{\Xi^{*}}}{k_1^2-m^2_{\Xi^{*}}+im_{\Xi^{*}}\Gamma_{\Xi^{*}}}[-g^{\nu\eta}+\frac{1}{3}\gamma_{\nu}\gamma_{\eta}+\frac{2}{3m^2_{\Xi^{*}}}k_{1\nu}k_{1\eta}+\frac{1}{3m_{\Xi^{*}}}(\gamma^{\nu}k_{1\eta}-\gamma^{\eta}k_{1\nu})],
\end{align}
with $q=k_2-p_1=p_2-k_1$, and $q^{'}=k_2-p_2=p_1-k_1$. We take $\Gamma_{\Xi^{*0}} = 9.1$ MeV, $\Gamma_{\Xi^{*-}} = 9.9$ MeV, $\Gamma_{\rho}= 149.1$ MeV, $\Gamma_{\omega} = 8.5$ MeV, $\Gamma_{\eta}=1.3$ MeV, $\Gamma_{\phi}=4.2$ MeV, and $\Gamma_{K^*} = 50.3$ MeV.  For the three-body final states $K^{-}\Xi^{0}\pi^{0}$ and  $\bar{K}^{0}\Xi^{-}\pi^{0}$, the isospin factors are $f_I=1$ and $-1$, respectively.  The isospin factor is $f_I=\sqrt{2}$ for all the other three-body final states.   To take into account the finite size of hadrons, for each vertex, we multiply a form factor $F(k_1^2)$ of the following form~\cite{Lin:2018nqd}
\begin{align}
F(k_1^2)=\frac{\Lambda^4}{(m^2-k_1^2)^2+\Lambda^4}\label{eqff},
\end{align}
where $m$ is the mass of the exchanged particle and $k_1$ is its momentum, with the cutoff $\Lambda$ varying from 0.8 GeV to 2.0 GeV.

In order to avoid  ultraviolet divergence in the triangle diagrams,  we take the three momentum truncation method to  compute the amplitudes.
In the rest frame of the $\Omega^*$, the relevant momenta are defined as follows:
\begin{align}
k_0&=(M,0,0,0),~~~~p_1=(E_{\Xi},0,0,p_{cm}),~~~~~p_2=(E_{K},0,0,-p_{cm}),\\
q&=(q_0,q_1\sin\theta,0,q_1\cos\theta)=q^{'},
\end{align}
and we can rewrite the $\int{}d^4q^{(')}=\int{}dq_0q_1^2dq_1d\cos{}\theta{}d\phi$ with $q_0 \epsilon{}(-\infty,\infty)$, $q_1\epsilon{}(0,\Lambda)$, $\cos\theta\epsilon(-1,1)$, and $\phi\epsilon{}(0,2\pi)$, where $\Lambda$ is a free parameter and is also taken to vary from 0.8 GeV to 2.0 GeV.

The partial decay width of the $\Omega^{*} \to \bar{K} \Xi$ and $\Omega^{*} \to \bar{K} \Xi\pi$ in its rest frame are given by
\begin{eqnarray}
d\Gamma_{\Omega^{*} \to \bar{K} \Xi}    & = & \frac{1}{32\pi^2} \overline{|{\cal M}|^2}  \frac{p}{M^2}d\Omega,\\
d\Gamma_{\Omega^{*}\to \bar{K} \Xi\pi} & = & \frac{1}{(2\pi)^5}\frac{1}{16M^2} \overline{|{\cal M}|^2} |\vec{p}_3^{*}||\vec{k}_2|dm_{\pi\Xi}d\Omega^{*}_{p_3}d\Omega_{k_2},
\end{eqnarray}
where $M$ is the mass of the $\Omega(2012)$, while $p$ is the module of the $\Xi$ (or $\bar{K}$) three-momentum in the rest frame of the $\Omega(2012)$. The $(|\vec{p}_3^{*}|,\Omega_{p_3}^{*})$ is the momentum and angle of the particle $\Xi$ in the rest frame of $\Xi$ and $\pi$, and $\Omega_{k_2}$ is the
angle of the $\bar{K}$ in the rest frame of the decaying particle.  The $m_{\pi\Xi}$ is the invariant mass for $\pi$ and $\Xi$ and $m_{\pi}+m_{\Xi}\leq{}m_{\pi\Xi}\leq{}M-m_{\bar{K}}$.
The averaged squared amplitude is then
\begin{eqnarray}
 \overline{|{\cal M}|^2}  &=& \frac{1}{4} \sum_{s_{\Omega^{*}}} \sum_{s_\Xi} |{\cal M}|^2,
\end{eqnarray}
where
\begin{eqnarray}
  {\cal M}_{\Omega^{*} \to \bar{K} \Xi} &=& {\cal M}_{\Sigma/\Lambda} + {\cal M}_{\rho/\omega/\phi} + {\cal M}_{K^*},\\
  |{\cal M}|^2_{\Omega^{*} \to \bar{K} \Xi\pi}&=&|{\cal M}|^2_{K^{-} \Xi^{0}\pi^{0}}+|{\cal M}|^2_{K^{-} \Xi^{-}\pi^{+}}+|{\cal M}|^2_{\bar{K}^{0} \Xi^{0}\pi^{-}}+|{\cal M}|^2_{\bar{K}^{0} \Xi^{-}\pi^{0}}.
\end{eqnarray}

\section{Numerical results and discussions}

As explained in the previous section, the triangle diagrams are regularized with a sharp momentum cutoff $\Lambda$, which is taken to be the same as those appearing in the form factor, $F(k_1^2)$.  Because the triangle diagrams are
ultraviolet divergent, our two-body decay width will depend on the value of the cutoff. Therefore,  it  is important to check whether one can obtain a decay width consistent with the experimental data using a reasonable value for the cutoff.

\begin{figure}[htbp]
\begin{center}
\includegraphics[scale=0.4]{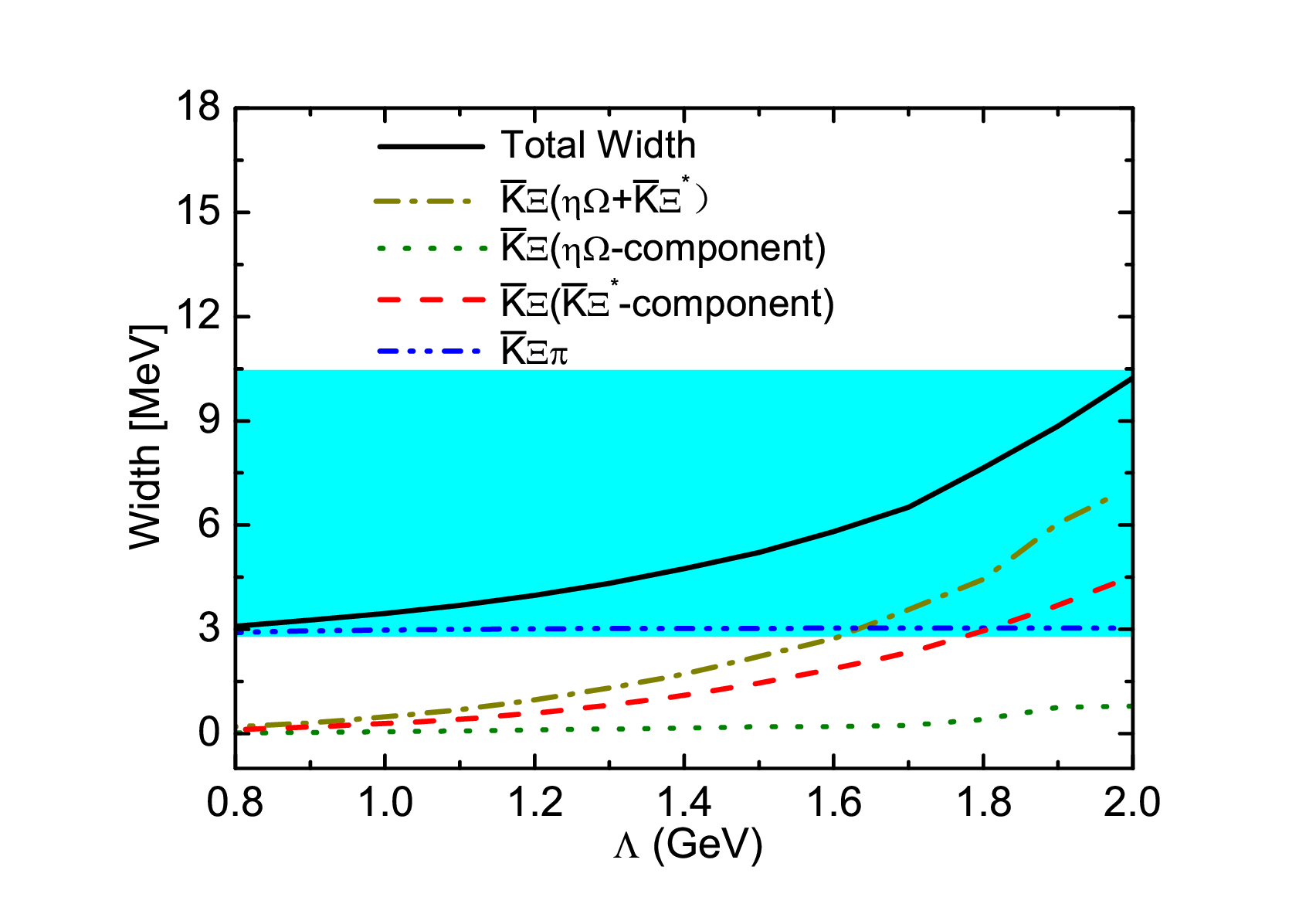}
\caption{(Color online) Total decay width of the $\Omega(2012)$ as a function of the parameter $\Lambda$. The cyan error bands correspond to the experimental decay width~\cite{Yelton:2018mag}.}
\label{t-mass}
\end{center}
\end{figure}

In Fig.~\ref{t-mass} we show the total decay width of $\Omega(2012) \to \bar{K} \Xi$ as a function of the cutoff parameter $\Lambda$. Note that the $\Omega(2012) \to \bar{K} \pi\Xi$ three-body decay does not depend on the cut-off parameter $\Lambda$, but it  depends weakly on the parameter $\Lambda_0$ as  in Ref.~\cite{Lin:2018nqd}.  We can see that the $\bar{K}\Xi(1530)$ component provides the dominant contribution to the partial decay width of the $\bar{K}\Xi$ two-body channel.   The $\eta\Omega$ contribution to the $\bar{K}\Xi$ two-body channel is very small.  However, the interference between them is still sizable and increases with the cutoff parameter $\Lambda$.

In Refs~\cite{Lin:2018nqd,Polyakov:2018mow},  the three-body decay was emphasized,  while  our result shows that two-body $\bar{K} \Xi$ decay width can become larger than the $\bar{K}\pi \Xi$ three-body decay width when $\Lambda$ is larger than $1.65$ GeV.  We note that
the  hyperon exchange and $\eta\Omega$ component contribution are not considered in   Ref.~\cite{Lin:2018nqd}.   More specifically, our three-body partial decay width, $\sim3.0$ MeV,  is close to the estimate of Ref.~\cite{Lin:2018nqd}, but smaller than that of Ref.~\cite{Polyakov:2018mow}, about10 MeV.  We note that our total decay width and that of
Ref.~\cite{Valderrama:2018bmv} agree with the experimental data.   The difference is  that Ref.~\cite{Valderrama:2018bmv} assumes that the $\Omega^{*}(2012)$ is a pure
$\Xi(1530)\bar{K}$ molecule and invokes some power counting arguments to calculate its two-body decay width.  Indeed, our study shows that the contribution  from the $\eta\Omega$ component is small.   Note that the molecule picture  is different from the $qqq$ picture of the  chiral quark model~\cite{Xiao:2018pwe} and light cone QCD sum rules~\cite{Aliev:2018syi,Aliev:2018}.

\begin{figure}[htbp]
\begin{center}
\includegraphics[scale=0.4]{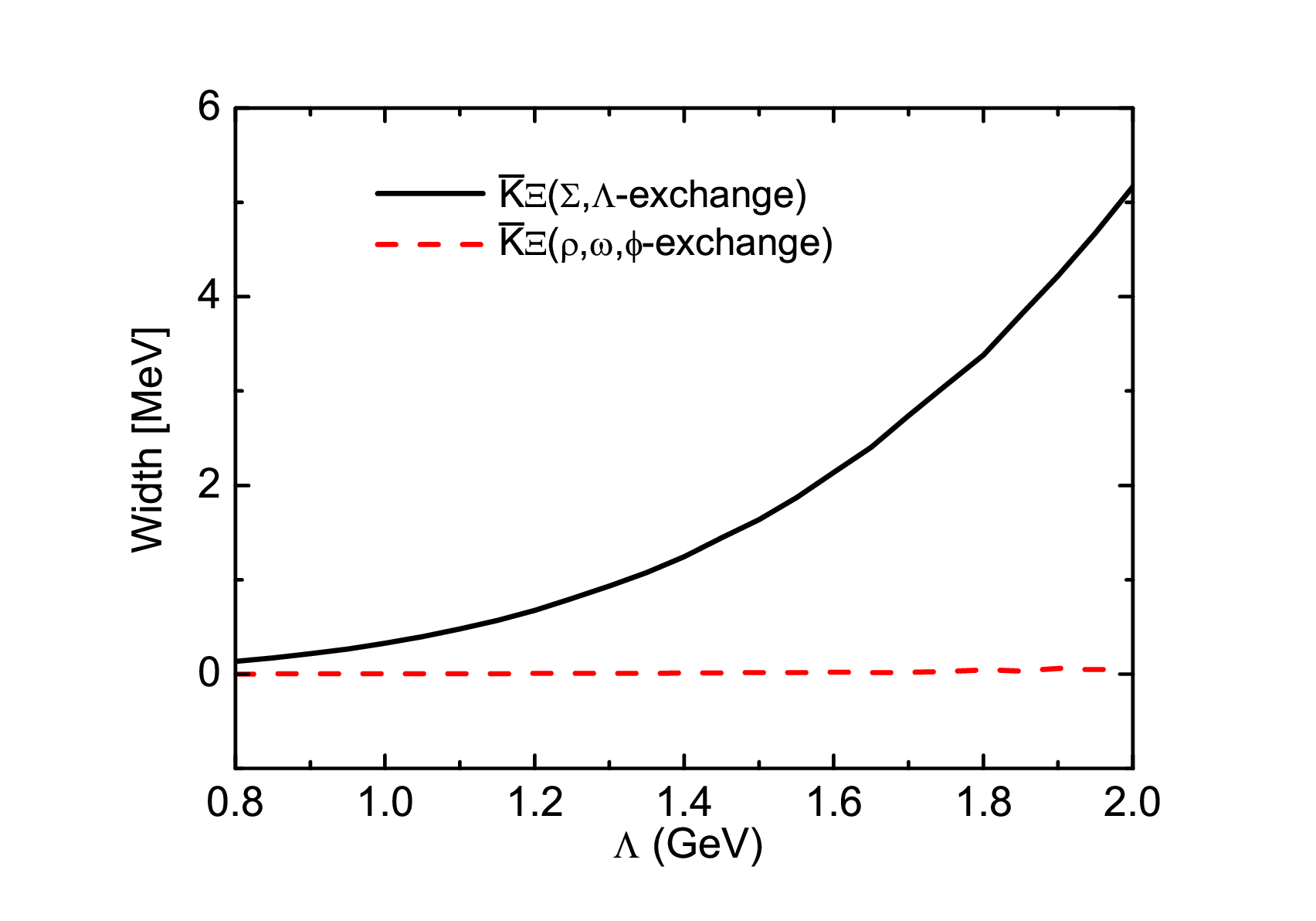}
\caption{(Color online) Decomposed contributions to the decay width of the $\Omega(2012)$ into $K\Xi$ as a function of the parameter $\Lambda$ }
\label{fig3}
\end{center}
\end{figure}

The contribution of the $\bar{K}\Xi(1530)$ component includes
two parts: i) $\Sigma$ and $\Lambda$ exchanges (fig.~\ref{mku} (a) and (b)); ii) $\rho$, $\phi$, and $\omega$ meson exchanges
(fig.~\ref{mku} (c) and (d)). The relative importance of these two mechanisms
to the $\Omega^* \to \bar{K} \Xi$ decay is shown
in Fig.~\ref{fig3}. One can see that the contribution
from $\Sigma$ and $\Lambda$ exchanges is larger than those from
$\rho$, $\phi$, and $\omega$ meson exchanges for the cutoff range studied.

\section{summary}
In summary, we have studied the $\bar{K} \Xi$ decay of the newly observed $\Omega^{*}(2012)$ assuming that the $\Omega(2012)$ is a dynamically generated state with
spin-parity $3/2^-$ from the coupled channel interactions of $\bar{K}\Xi(1530)$ and $\eta \Omega$ in $S$-wave.
Take $\alpha_{\mu}=-2.5$ and $\mu=0.7$ GeV,  we obtained a pole at $M=2012.7$ MeV, and associated it
to the newly observed $\Omega^{*}$(2012).   With the coupling constants between the  $\Omega^{*}(2012)$ and its components calculated from the residue at the pole position, $g_{\Omega^{*}\bar{K}\Xi^{*}}$=1.65 and $g_{\Omega^{*}\eta\Omega}$=2.80, we obtained the partial decay widths of the $\bar{K}\Xi$ final
state through triangle diagrams in an effective Lagrangian approach.   In such a picture, the decay $\Omega^{*}(2012)\to\bar{K}\Xi$
occurs by exchanging $\Sigma,\Lambda$ hadrons and $\rho,\phi,\omega$,and $K^{*}$ vector mesons.  The contribution to the three-body decay $\Omega(2012) \to \bar{K} \Xi \pi$
only comes from the $\bar{K}\Xi(1530)$ component.

 It was shown that the calculated total decay width of the $\Omega^{*}(2012)$ is in fair agreement with the experimental data, thus supporting the assignment of its spin-parity as $3/2^-$, In addition, we showed that the $\eta\Omega$ channel plays a less relevant role.

The present work should be viewed as a natural extension of the works of Refs.~\cite{Si-Qi:2016gmh,Sarkar:2004jh}, where the chiral unitary approach
was employed to dynamically generate such an exited $\Omega^*$. The present work showed indeed
that the chiral unitary approach can provide a satisfactory explanation of not only the mass but also the decay width
of the $\Omega(2012)$, consistent with the experimental data.

\section*{Acknowledgements}

JJX and LSG thank Qiang Zhao and Bing-Song Zou for valuable discussions. This work is partly supported by the National Natural
Science Foundation of China under Grants No. 11522539,
No.1 1735003, and No. 11475227, the fundamental Research Funds for the Central
Universities and the Youth Innovation Promotion Association
CAS No. 2016367.


\end{document}